\documentclass[11pt]{article}
\usepackage{graphicx}
\usepackage{amssymb}
\usepackage{epstopdf}
\usepackage{amsmath,amsfonts}

\newtheorem{theorem}{Theorem}
\newtheorem{corollary}[theorem]{Corollary}
\newtheorem{lemma}[theorem]{Lemma}

\newtheorem{remark}{Remark}
\begin{document}
\title{Maxwell's equations}
\author{Daniel Henry Gottlieb}
\maketitle
\begin{abstract}
We express Maxwell's equations as a single equation, first using the divergence of a special
type of matrix field to obtain the four current, and then the divergence of a special matrix to obtain
the Electromagnetic field. These two equations give rise to a remarkable dual set of equations
in which the operators become the matrices and the vectors become the fields. The decoupling
of the equations into the wave equation is very simple and natural. The divergence of the stress
energy tensor gives the Lorentz Law in a very natural way. We compare this approach to the
related descriptions of Maxwell's equations by biquaternions and Clifford algebras.

\end{abstract}

\section{Introduction}

Maxwell's equations have been expressed in many forms in the century and a half
since their discovery. The original equations were 16  in number. The vector forms,
 written below, consist of 4 equations. The differential form versions consists of two
 equations; see [Misner, Thorne and Wheeler(1973); see equations 4.10, 4.11]. See also
 [Steven Parrott(1987) page 98 -100 ]
 The application of quaternions, and their complexification, the biquaternions,
 results in a version of one equation. William E. Baylis (1999) equation 3.8, is an example. 
 
 In this work, we obtain one Maxwell equation, (10), representing the electromagnetic 
 field as a matrix
 and the divergence as a vector multiplying the field matrix. But we also obtain a remarkable
 dual formulation of Maxwell's equation, (15), wherein the operator is now the matrix and 
 the field is now the vector. The relation of the four vector potential 
 to the electromagnetic field has the same sort of duality; see equations (13) and (14).
 
 These dual pairs of equations are proved equivalent by expanding the matrix multplication
 and checking the equality. Indeed, they were discovered this way. However, it is possible
 to ask if there is an explanation for this algebraic miracle. 
 There is. It follows from the commutivity of certain types of matrices. This commutivity
 also implies the famous result that the 
 divergence of the stress-energy tensor  $T$  is the electrom magnetic field $F$ applied to the the
 current-density vector. 
 
 The commutivity itself, again discovered and proved by a brute force calculation, has an 
 explanation arising from two natural representations of the biquaternions; namely the left
 and the right regular representations.
 
 Why do matrices produce an interesting description  of Maxwell's equations, when tensors
 are so much more flexible? It is because matrices are representations of linear transformations
 for a given choice of a basis. The basis is useful for calculation, but reformulating concepts 
 and definitions in terms
 of the appropriate morphisms (in this case, linear transformations) almost always pays a 
 dividend, as we topologists have discovered during the last century.

This paper arose out of an email from me to  Vladimir Onoochin concerning questions about
Maxwell's equations. I thank him for a very interesting correspondence.

\section{Maxwell's Equations}

We are not really concerned with physical units for our paper, however what we have
written is compatible with natural Heavyside-Lorentz units where 
the speed of light $c$ = 1 and the electric permittivity $\epsilon_0 = 1$ .
See Baylis(1999), section 1.1.

We will say that two vector fields ${\bf  E}$ and  ${\bf  B}$ satisfies Maxwell's equations if

\begin{eqnarray}
\nabla \times \bf E + \partial_t  \mathbf {B} & =  & \mathbf{0}\\
\nabla \times \bf B  - \partial_t  \mathbf {E} & =: & \bf J \\
\nabla \cdot \bf E & =: & \rho \\
\nabla \cdot \bf B  & =  & 0
\end{eqnarray}

\noindent Here $=:$ means the right symbol is defined by the left side of the equation. 
Now we follow [Gottlieb (1998)] and [Gottlieb (2001)] and recall the notation
for Lorentz transformations.  Let $M$ be Minkowski space with inner product
$\langle \ , \ \rangle$ of the form $- + + +$.  Let $e_0, e_1, e_2, e_3$ be an orthonormal basis
with $e_0$ a time-like vector.  A linear operator $F: M \rightarrow M$ which
is skew symmetric with respect to the inner product $\langle \ , \ \rangle$ has a matrix 
representation, depending on the orthonormal basis, of the form 
\begin{equation}
 F = \left(
\begin{array}{cc}
 0&   \vec E^t \\
 \vec E& \times \vec B
\end{array}
\right)
\label{F}
\end{equation}

\noindent where $\times \vec B$ is a $3 \times 3$ matrix such that $(\times \vec B) \vec v = 
\vec v \times \vec B$, the cross product of $\vec v$ with $\vec B$.  That is
\[
\left(
\begin{array}{ccc}
  0&B_3   & -B_2  \\
  -B_3 & 0  &B_1   \\
B_2  & -B_1  &0   
\end{array}
\right)
\]
where $\bf B$ is given by $(B_1, B_2, B_3 )$.

The dual $F^*$ of $F$ is given by 

\begin{equation}
F^* := \left(
\begin{array}{cc}
   0& - \vec B^t\\
       -\vec B& \times \vec E\\
\end{array}
\right)
\end{equation}
  
We complexify $F$ by $cF:= F - i F^*$.  Its matrix representations is
\begin{equation}
   cF= \left(
   \begin{array}{cc}
     0& \vec A^t\\
               \vec A& \times (-i \vec A)\\ 
            \end{array}
           \right)
        \hbox{where } \vec A = \vec E + i \vec B
\end{equation}

\begin{remark}
These complexified operators satisfy some remarkable Properties:

\begin{description}
\item[a)] $cF_1 cF_2 + cF_2 cF_1 = 2 \langle \vec A_1, \vec A_2 \rangle I$ where $\langle \ , \ \rangle$
denotes the complexification of the usual inner product of $\mathbb R ^3$.
Note, $\langle \ , \ \rangle$ is not the Hermitian form, that is our inner
product satisfies $i \langle \vec v, \vec w \rangle =
\langle i \vec v, \vec w \rangle = \langle \vec v, i \vec w \rangle$.

\item[b)] The same property holds for the complex conjugates $\overline cF_1$
and $\overline cF_2$. The two types of matrices commute. That is  $ cF_1 \overline cF_2
= \overline cF_2  cF_1$

\item[c)]$ c F \overline c F=: 2T_F$ where $T_F$ is proportional to the stress-energy 
tensor of electromagnetic fields.
\end{description}

Now $e^F = I + F + F^2 / 2!  +  F^3 / 3! + \ldots : M 
\rightarrow M$ is a proper Lorentz transformation.  It satisfies 
$e^F = e^{cF / 2} e^{\overline cF / 2 }$.  The algebraic properties of $c F$ give rise 
to a simple expression for its exponential:
$$
   e^{cF} = \cosh (\lambda_{cF})I + \frac{\sinh (\lambda_{cF})}{ \lambda_{cF}}   cF
$$
where $\lambda_{cF}$ is an eigenvalue for $cF$.  There are only, at most, two
values for the eigenvalues of $cF$, namely $\lambda_{cF}$ and $- \lambda_{cF}$.
\end{remark}

Now let us see how to describe Maxwell's field equations by means of these matrices.
Now consider the $4 \times 4$ matrix $F$ defined by equation (\ref{F}). We multiply it by the
$4 \times 1$ vector of operators $(-\partial_t , \nabla)^t$
to get
\begin{equation}
F{ -\partial_t  \choose \nabla} = {\rho \choose \bf J }
\label{current}
\end{equation}

\noindent and

\begin{equation}
F^*{- \partial_t  \choose \nabla} = {0 \choose \bf 0 }
\label{zero}
\end{equation}
See equation (23) for more details on this notation.

Equations \ref{current} and \ref{zero} hold if and only if Maxwell's equations (1 - 4) hold.
Using the definition of $cF$ in equation 
(7) we get

\begin{equation}
cF{ -\partial_t  \choose \nabla} = {\rho \choose \bf J }
\end{equation}

\noindent is true if and only if $E$ and $B$ satisfy Maxwell's equations.

In fact, the version of Maxwell's equations involving the four vector fields $\bf {E, \ \ B, \ \ D, \ \ H}$
is equivalent to the matrix equation

\begin{equation}
\left(
\begin{array}{cc}
  0&\bf {D + iB}    \\
\bf {D} + i \bf {B } & \times (-i(\bf {E} + i \bf {H}))    \\
 
\end{array}
\right){- \partial_t  \choose \nabla} = {\rho \choose \bf J }
\end{equation}

\begin{remark}

Note the forms of the complex matrices in equation (7) and in equation (11): respectively

\[
   \left(
   \begin{array}{cc}
     0 & \vec A^T\\
               \vec A& \times (\pm i \vec A)\\ 
            \end{array}
           \right) \textnormal{   and   }
   \left(
   \begin{array}{cc}
     0 & \vec A^T\\
               \vec A& \times (\vec C)\\ 
            \end{array}
           \right)
 \]
 
 Now within the class of all matrices of the  second form, only the first form has the property that
 its square is equal to a multiple of the identity.
 
\end{remark}

Now let us consider a four vector field ${\varphi \choose \bf A}$. We define an associated
$\bf E $ and $\bf B$ by the equations
\begin{equation}
\nabla \times \bf A =: \bf B \textnormal{   and   } -\partial_t  \bf A - \nabla \varphi =: \bf E
\end{equation}

Then the $\bf E $ field and the $\bf B$ field defined above satisfy Maxwell's equations (1 - 4). 
We can describe the four vector field by means of a similar matrix equation. Let $I$ denote
the $4 \times 4$ identity matrix. 

\begin{equation}
\left(\varphi I + 
\left(
\begin{array}{cc}
0  & \bf A^t    \\
 \bf A & -i \times \bf A    \\
     
\end{array}
\right)\right) {\partial_t \choose \nabla} = {\partial_t \varphi + \nabla \cdot \bf A \choose -\bf E -i\bf B}
\end{equation}

Now a remarkable duality holds. The following equation is also equivalent to the above equation.

\begin{equation}
\left(\partial_t I + 
\left(
\begin{array}{cc}
0  & \bf \nabla^t   \\
 \bf  \nabla& i \times \bf \nabla   \\
  
\end{array}
\right)\right) {\varphi \choose \bf A} = {\partial_t \varphi + \nabla \cdot \bf A \choose -\bf E -i\bf B}
\end{equation}

Similarly, equation (10) has a dual equation which holds if and only if Maxwell's equations are 
satisfied,

\begin{equation}
\left(\partial_t I -
\left(
\begin{array}{cc}
0  & \bf \nabla^t    \\
 \bf  \nabla& i \times \bf \nabla   \\
 
\end{array}
\right)\right) {0 \choose -\bf E -i\bf B} = {\rho \choose \bf J}
\end{equation}

These equations give rise to an interesting question:

\begin{remark}

Can we give an explanation of the remarkable dualities between equations
(10) and (15) and between equations (13) and (14) ? Yes! It is based on the fact that
matrices of the form $cF$ commute with the matrices of the form $\overline cF$. Hence
the matrix on the left hand side of (13) commutes with the matrix on the left hand side of
of (14). Thus the first columns of two products of the 
commuting matrix products must be equal. Since the
vectors on the left hand side of (13) and 14) are the first columns of each of the two product matrices
that commute, it follows that their products are equal. Exactly the same argument holds for
equations (10) and (15).

\end{remark}

Now from part  a) of {\bf Remark} 1, we know that 

\begin{equation}
\left(
\begin{array}{cc}
0  & \bf \nabla^t    \\
 \bf  \nabla& i \times \bf \nabla   \\
 
\end{array}
\right)^2 = \nabla^2 I
\end{equation}

Consistent with our notation we define
\begin{equation}
\left(
\begin{array}{cc}
0  & \bf \nabla^t     \\
 \bf  \nabla& i \times \bf \nabla   \\
   
\end{array}
\right) =:  \overline c \nabla
\end{equation}
so that
\begin{equation}
(\partial_t I - \overline c \nabla)(\partial_t I + \overline c \nabla) =  (\partial_t I + \overline c \nabla)(\partial_t I - \overline c \nabla) = (\partial_t^2 - \nabla^2) I 
\end{equation}

Now apply $ (\partial_t I + \overline c \nabla)$ to equation (15) and obtain the wave equation
and obtain the wave equation
\begin{equation}
(\partial_t^2 - \nabla^2){0 \choose -\bf E -i\bf B} = \nabla \rho + \partial_t J -i \nabla  \times J
\end{equation}
implying the conservation of charge equation $ \partial_t  \rho + \nabla \cdot \bf J = 0$

On the other hand, if we apply $( \partial_t I - \overline c \nabla )$ to equation (14) we obtain,
thanks to equation (15),
\begin{equation}
(\partial_t^2 - \nabla^2) {\varphi \choose \bf A}  = {\rho  \choose \bf J} + \textnormal {vector depending on    } \partial_t \varphi + \nabla \cdot \bf A
\end{equation}

So if the covariant gauge condition is chosen, $\partial_t \varphi + \nabla \cdot \bf A = 0$,
we have the wave equation  
\begin{equation}
(\partial_t^2 - \nabla^2) {\varphi \choose \bf A}  = {\rho  \choose \bf J} 
\end{equation}

\section{The electromagnetic stress-energy tensor}

Not only is equation (10) equivalent to Maxwell's equations, but its complex conjugate below
is also equivalent to Maxwell's equations, since the current is real.

\begin{equation}
\overline cF{ -\partial_t  \choose \nabla} = {\rho \choose \bf J }
\end{equation}

Thus we have equations (10), its complex conjugate (22), and its dual (15), and the complex
conjugate of (15) all being equivalent to the vector form of Maxwell's equations (1--4). Thus we
not only reduce the number of Maxwell's equations from 4 to 1, we obtain 4 equivalent equations.
Thus, by a talmudic argument, we can say we have reduced the four Maxwell's equation to
1/4 of an equation. These different forms of the equation interact with each other to produce
new derivations of important results.

We have used the following notation of the divergence of a matrix field. What we mean by the notation

\begin{equation}
A{ -\partial_t  \choose \nabla} 
\end{equation}

is that the column vector of operators multiplies into the matrix $A$ of functions
and then the operators are applied to the functions they are next to. In index notation we obtain a 
vector whose $i$--th row is  ${a_i}_j\partial_j := \partial_j( {a_i}_j )$.

Another way to achieve the same result is to take the differential of the matrix,
$dA$. Here we obtain a matrix of differential one-forms whose $(i,j)$-th element is

\begin{equation}
d{a_i}_j = \partial_k({{a_i}_j) }dx_k
\end{equation}
 
Then we employ the differential geometry convention that $dx_i(\partial_j)=\delta_i{_j}$.
Thus $d{a_i}_j (\partial_j)=\partial_j({a_i}_j) $. Thus our definitions leads us to

\begin{equation}
A{ -\partial_t  \choose \nabla} = dA\cdot { -\partial_t  \choose \nabla}
\end{equation}

Now the Leibniz rule gives us

\begin{equation}
d(AB) = (dA)B +A(dB)
\end{equation}

This equation helps us to study the divergence of a matrix product. In particular,
if $dA$ and $B$ commute, then the divergence of the product is B times the divergence
of A  plus A times the divergence of B.

\begin{theorem}
Let $T_F$ be the electromagentic stress energy tensor of the electromagnetic matrix
field $F$. Then 
\begin{equation}
T_F { -\partial_t  \choose \nabla} = F{ \rho  \choose \bf J} 
\end{equation}

\end{theorem}

Proof: 

The first equation below follows from (25). The second equation follows from {\bf Remark 1 c)}.
The third equation follows from (26). The fourth equation follows from {\bf Remark 1 b)}, which is
the commutivity of matrices of the form $cF$ with matrices of the form $\overline cG$, combined 
with the observation that $d(cF)$ has the same form as $cF$ and similarly $d(\overline cF)$ has
the same form as $\overline cF$. The fifth equation follows by linearity . The sixth equation follows
from the form of Maxwell's equations found in equation (10) and its complex conjugate equation (22).
Finally the seventh equation follows from the definition of $cF$ and the fact that $F$ is its real
part.
$$ T_F { -\partial_t  \choose \nabla}= d(T_F)\cdot { -\partial_t  \choose \nabla} ={ 1 \over 2 }
d(cF \overline cF) \cdot  { -\partial_t  \choose \nabla} =  { 1 \over 2 }(d(cF) \overline cF + 
cF d(\overline cF))\cdot  { -\partial_t  \choose \nabla} = $$
$$ { 1 \over 2 }(cF  d(\overline cF) + cF d(\overline cF))\cdot  { -\partial_t  \choose \nabla}
=  { 1 \over 2 }cF  d(\overline cF)\cdot  { -\partial_t  \choose \nabla}  +
{ 1 \over 2 } \overline cF  d(cF)\cdot  { -\partial_t  \choose \nabla} = $$
$${ 1 \over 2 }cF  { \rho \choose \bf J}  +
{ 1 \over 2 } \overline cF    {\rho \choose \bf J} = F{ \rho \choose \bf J}  $$
$\square$

\section{Biquaternions}

We choose an orthogonal coordinate system $(t,x,y,z)$ for Minkowski space $\Bbb R^{3,1}$.
Let $e_i$ be the corrosponding unit vectors (with respect to the Minkowski metric).
We express $\vec e_1=(1,0,0)^t,\ \vec e_2=(0,1,0)^t,\ \vec e_3=(0,0,1)^t$. So our standard
choice of basis  is given by $e_i = (0, \vec e_i)^t$ for $i \neq 0$.

Let $cE_i$ be the matrix below where $\vec A = \vec e_i$ where $i=1,2,3$.

\begin{equation}
   cE_i= \left(
   \begin{array}{cc}
     0& \vec A^T\\
               \vec A& \times (-i \vec A)\\ 
            \end{array}
           \right)
        \hbox{where } \vec A = \vec E + i \vec B = \vec e_i
\end{equation}

\begin{remark} a) The set of sixteen matrices
$$
I, \ \ cE_i,\ \ \overline cE_i, \ \ cE_i \overline cE_i
$$
where $i = 1, 2, 3$,
forms a basis for $M_4(\Bbb C)$, the vector space of $4\times 4$ complex matrices.

b) The square of each matrix in the basis is I.

c) Each matrix is Hermitian, and all the matrices, except for the identity $I$, have zero trace.

d) $cE_1cE_2 = icE_3$ and $\overline cE_1\overline cE_2 = -i\overline cE_3$.

e) $cE_i$ and $ cE_j$ anti commute when $i \neq j$; and 
$\overline cE_i$ and $ \overline cE_j$ anti commute when $i \neq j$. 
 Also $cE_i$ and $\overline cE_j$
commute for all $i$ and  $j$.
\end{remark}

See Gottlieb(2001), Theorem 3.3.

Now from the above theorem, we see that the 4 elements $\{ I, cE_i \}$ form a basis for the
biquaternions. Baylis(1999) chooses to denote his basis for the biquaternions, which he views
as the real Clifford algebra on three generators, with a basis $\bf e_i$ where $i=0,1,2,3$ where
$\bf e_0$ is the multiplicative identity and $\bf e_i^2=1$, and the $\bf e_i$ anti commute for 
$i\neq 0$. Baylis chooses the orientation $\bf e_1\bf e_2= i \bf e_3$. 
Thus our representation of Baylis's biquaternions is given by ${\bf e_0} \mapsto I$ and
${\bf e}_i \mapsto cE_i$.

Let us denote the biquaternions given above by $ P$. The alternative choice where
the biquaternions satisfy the relation $\bf e_1\bf e_2= - i \bf e_3$ will be denoted $\overline P$ and is
represented by ${\bf e}_i \mapsto \overline cE_i$. The symbol $P$ stands for paravectors, which is Baylis'
name for the space of biquaternions. The name paravectors stems from Baylis' underlying
Clifford algebra approach.

Given a paravector $A \in P$, we define { \it left multiplication by $A$  } as

$$L_A: P \rightarrow P: X \mapsto AX$$ and { \it right multiplication by $A$  } as
$$R_A: P \rightarrow P: X \mapsto XA$$

\begin{lemma}
Both left multiplication and right multiplication by $A$ are linear transformations,
and they commute: that is $L_AR_B = R_BL_A$
\end{lemma}

Now given the basis $ \{ \bf e_i \}$, the linear transformations can be represented by
matrices muliplying the coordinate vectors from the left.

\begin{theorem}
$L_A $ is represented by the matrix $aI + cF$ corresponding to $A$. Also, 
$R_A$ is represented by the transposed matrix $(aI +cF)^t$. If $A \in \overline P$, then $A$ is 
represented by $aI + \overline cF$ and so $R_A$ is represented by the transposed
 $(aI + \overline cF)^t$.
\end{theorem}
proof: Consider $L_1$, by which we mean left multiplication by $\bf e_1$. So the basis elements
are transformed by $L_1$: 
$\bf e_0 \mapsto e_1e_0 = e_1$,
$\bf e_0 \mapsto e_1e_1 = e_0$,
$\bf e_0 \mapsto e_1e_2 = ie_3$,
$\bf e_0 \mapsto e_1e_3 = -ie_2$.
This corresponds to the matrix

\[
\left(
\begin{array}{cccc}
  0& 1   & 0   & 0 \\
  1 & 0  & 0 & 0   \\
   0 & 0  &0   & -i   \\
   0 & 0  & i  & 0   \\

\end{array}
\right)
\]
This matrix is $cE_1$. Right multiplication by $\bf e_1$ gives us the matrix $\overline cE_1$.
Note that this is the transpose of $cE_1$. In the same way we see that
left multiplication by $\bf e_i$ gives rise to the matrix $cE_i$ and right multiplication by
$\bf e_i$ gives rise to $cE_i^t = \overline cE_i$.

Now every element $A \in P$ is a unique linear combination  $a_ie_i$, so the representation
of $A$ is the same linear combination of the $cE_i$ : Namily, $a_icE_i$. Thus $L_A$ is represented 
by $ a_icE_i$. Also $R_A$ is represented by $(a_icE_i)^t= a_i \overline cE_i$
$\square$

Baylis(1999) in his textbook, describes an algebraic system in which the paravectors
are uniquely written as linear combinations $a_i\bf e_i$. The choice of the basis
implicitly defines an isomorphism

\begin{equation}
\Phi: P \rightarrow {\mathbb C}^4: X = a_i{\bf e_i}\mapsto (a_0, a_1, a_2, a_3)^t
\end{equation}

Whereas if $v$ is a nonnull vector in complex Minkowski space $M$ the evaluation map
is an isomorphism 

\begin{equation}
\Theta_v: P \rightarrow {\mathbb C}^4: X = a_i{cE_i}\mapsto Xv
\end{equation}
proof:
See Gottlieb(1999), Theorem 6.9.

Note that if $v = {e_0}$, then $\Phi = \Theta_ {\bf e_0}$. This follows since
$a_icE_i\mapsto a_icE_ie_0 = a_ie_i =  (a_0, a_1, a_2, a_3)^t$.

Now Baylis(1999) exposes Electromagnetism in the language of Clifford algebra
on three generators, which is isomorphic to the biquaternions. We therefore want 
to adopt his notation as much as possible, while thinking of our matrices replacing
his abstract symbols. Our goal is not to advance a new method of calculation so much 
as to use the matrices to understand the underlying geometry which arises out of an
electromagnetic field. Our point of departure, explained in Gottlieb(1999), is to consider
the linear transformation $F: M \rightarrow M$ which arises in the Lorentz law. Choosing an orthonormal basis
$\{e_i\}$ with respect to the Minkowski metric $< , >$, we consider those linear transformations
$F: M \rightarrow M$ which are anti symmetric with respect to the Minkowski metric. That is

\begin{equation}
\overline F = -F 
\end{equation}
where  $\overline F$ is defined by the equation 

\begin{equation}
<\overline Fv,w> = < v, Fw>
\end{equation}
That is, $\overline F$ is the adjoint of $F$ with respect to the Minkowski metric.

The matrix form
which these skew symmetric operators take is that of the matrix in equation (5). Now our point of view is 
the following. The Minkowski metric is taken to be the primary object, based on the work of Robb(1936),
who produced a protocol as to how the metric could be measured by means of light rays. Also we 
take linear transformations as primary, since they are the morphisms of the category of
vector spaces and linear transformations. This point of view is suggested by the successes
of algebraic topology. We do not hesitate to employ other inner products on $M$ or the convenience
of tensors, but we will never change the underlying sign of the metric to match the sign suggested
by the wave equation.

For example, the Minkowski metric $< , >$ is taken to be defined independently of a choice of basis.
Having chosen an orthonormal basis, we of course represent $F$ by a matrix of type (5). We can
also define based on this choice of a basis, the Euclidean metric $< , >_C$ and the Hermitian
metric $< , >_H$. Now the adjoints of $F$ are defined by $< F^tv , w>_C = < v , Fw>_C$ and
$< F^\dagger v, w>_H = < v, Fw>_H$. On the matrix level, $t$ is the transpose and $\dagger$ is
the complex conjugate transpose. The adjoint ${\overline {(aI + cF)}} = aI -cF$.

Now Clifford algebras have certain involutions which are defined ad hoc and are very
useful in Clifford algebras. These are { \it Clifford conjugation (or spatial reversal)} denoted by
$p \mapsto \overline p$ and { \it hermitian conjugation (or reversion)} denoted by
$p \mapsto p^\dagger$

Now starting from a real linear transformation, the definition of $cF$ is seems just a clever
trick that is very useful. If you take a point of view that the biquaternions are the key concept,
then the complexification trick is explained. Similarly, the Clifford conjugation and hermitian
conjugation are tricks from the point of view of Clifford algebra, but from the matrix point of
view they correspond to the Minkowski adjoint and the Hermitian adjoint respectively.

Another important involution is {\it complex conjugation} denoted by  $p \mapsto p^{\overline c}$
The bar is reserved for Clifford conjugation, and the * is reserved for the 	Hodge dual , see equation (6).
Now the space of $4\times 4$ matrices is a tensor product of $P$ and $\overline P$. Now 
transpose $t: P \rightarrow \overline P$ and complex conjugation $\overline c:P \rightarrow \overline P$.
The composition of complex conjugation and transpose is the Hermitian adjoint 
$\dagger : P \rightarrow P$. Both the Minkowski adjoint and the Hermitian adjoints reverse the
orders of multiplication because they are both adjoints. 
Thus $\overline {AB} = \overline {B} { \ \  \overline A}$ and $(AB)^{\dagger} = B^{\dagger} A^{\dagger}$.
Complex conjugation is important on the underlying vector space $\mathbb C^4$ where it is given
in our new notation by $(a_i e_i)^{\overline c} = a_i^{\overline c}e_i$. Using the isomorphism $\Phi$,
see (29),
it can be grafted onto the biquaternions as $(a_i {\bf e_i})^{\overline c} = a_i^{\overline c} \bf e_i$.
This definition is not used in Baylis(1999) probably because it has no simple relation to the
ring structure of the biquaternions. However, the natural extension of complex conjugate to matrices,
given by $(a_{ij})^{\overline c} = (a_{ij}^{\overline c})$, is important on the matrix level. For our
matrix description of biquaternions, we get 
\begin{equation}
(aI + cF)^{\overline c} = a^{\overline c}I + \overline cF
\end{equation}

If we apply transpose to this equation we get
\begin{equation}
(aI + cF)^{\dagger} = a^{\overline c}I + cF^\dagger
\end{equation}
The problem for a strictly biquaternion approach is that the right side of (33) is no longer in
the the biquaternions $P$, and although $\overline c$ does preserve order of products, it also is not
linear over the complex numbers.

The composition of Hermitian  conjugation with Clifford conjugation leads to an product order
preserving automorphism  
$ + : P \rightarrow P : (aI + cF) \mapsto (aI + cF) ^+ = a^{\overline c}I  - cF^\dagger$. This automorphism
is called the grade automorphism.

Our Hodge dual, see (6), eventually differs with that of Baylis. We have two choice for the Hodge
dual, either the matrix in (6) or its negative. With our definition $cF := F -iF*$ we agree with
Baylis' choice of orientation $e_1e_2 =ie_3$ via the left regular representation. 
Then extending the definition of dual to 
$cF$ we use $cF^* = (F -iF^*)^* := F^* -iF^{**} = F^* +iF = i(F -iF^*) = icF$. So in $P$ it looks
like the definition should be multiplication by $i$. If we used the alternative definition of 
$(\overline cF)^* = -i \overline cF^*$ this gives multiplication by $-i$. But $P$ agrees with the
orientation $e_1e_2e_3 = i$.  Baylis chooses to define the Hodge dual by 
$p^* := pe_3e_2e_1 $, which is effectively multiplying by $-i$.

\section{Biparavectors}

Let $cE_0 := I$, and let $\overline cE _0 := I$ Then $\{ cE_i \otimes  cE _j \}$ forms a vectorspace
basis for $P \otimes P$. Similarly, $\{ cE_i \otimes \overline cE _j \}$ forms a vectorspace basis for 
$P \otimes \overline P$. Now we will call any element in $P \otimes P$ a {\it biparavector}. So a biparavector
is a sum of 16 terms give by $a_{ij} cE_i \otimes  cE _j$. Now a bivector gives rise to a linear
transformation $T : P \rightarrow P :X  \mapsto a_{ij} cE_iX cE _j$ In fact, every linear transformation
can be represented this way as a biparavector. Now by {\bf Theorem 3} the linear transformation
given by the biparavector $\{ a_{ij}cE_i \otimes  cE _j \}$  is represented by the matrix
$\{a_{ij} cE_i \overline  cE _j \}$

\begin{theorem}
Let $A = ac_{ij} E_i \otimes  cE _j$ be a biparavector representing a linear transformation $M$,
of $ \mathbb C^4$, repersented by the matrix again called $M$.
\newline a) Then $M = a_{ij} cE_i \overline cE _j$ where $a_{ij} = \frac 1 4 trace (McE_i \overline cE_j)$. 
\newline b) Let $x \in \mathbb C^4$ correspond to $ X \in P$ by $Xe_0 = x$. 
Then $Mx = A(X)e_0 := a_{ij}c E_i X  cE _j e_0$
\end{theorem}
proof: follows from remark 4. The trace from part a) follows from {\bf Remark 4, c)}. The trace
of the product of two matrices $tr(AB) = tr(BA)$ gives rise to an innerproduct on the vector space of
matrices. The the above cited remark shows that the basis in question is orthnormal with respect
to the inner product given by $\frac 1 4 trace$.

\begin{corollary}

a) The stress-energy tensor $T_F = \frac 1 2cF\overline cF$, so the equivalent  biparavector is
$cF \otimes  cF^\dagger$
\newline b) A real proper Lorentz transformation 
$e^F = e^{{\frac 1 2}cF} e^{{\frac 1 2}\overline cF}$
So the equivalent  biparavector is $e^{{\frac 1 2}cF} \otimes e^{{\frac 1 2}cF^{\dagger}} $
\end{corollary}
proof:
The key point is that $\overline cF^t = cF^{\dagger}$


\noindent{ \bf REFERENCES}

\noindent William E. Baylis (1999); Electrodynamics: A Modern Geometric approach; Birkhauser, Boston

\par

\noindent Daniel H. Gottlieb (1998)
\noindent Skew Symmetric Bundle Maps 
\noindent Contemporary Mathematics
\noindent vol. 220
\noindent pp. 117 - 141

\par

\noindent Daniel H. Gottlieb (2001)
\noindent Fields of Lorentz transformations on space-time
\noindent Topology and its Applications
\noindent vol. 116
\noindent pp. 103 - 122

\noindent Misner, Thorne and Wheeler (1973)W.H Freeman and Company New york
 
 \noindent Stephen Parrott;(19870)Relativistic and Electordynamics and Differential 
Springe Verlag. New York

A. A. Robb (1936) Geometry of Time and Space. Cambridge University Press, 1936.

\end{document}